\documentclass[12pt,amsmath,amssymb]{iopart}

\usepackage{epsfig}

\begin{document}

\title{Contact processes with competitive dynamics
in bipartite lattices: Effects of  distinct interactions}
\author{Salete Pianegonda}
\address{Instituto de F\'{\i}sica, Universidade Federal 
do Rio Grande do Sul, Caixa Postal 15051, CEP 91501-970, Porto Alegre, RS, Brazil}
\author{Carlos E. Fiore}
\address{Instituto de F\'{\i}sica,
Universidade de S\~{a}o Paulo, 
Caixa Postal 66318\\
05315-970 S\~{a}o Paulo, S\~{a}o Paulo, Brazil}

\date{\today}

\begin{abstract}
The two-dimensional contact process (CP) with a competitive dynamics
proposed by Martins {\it et al.} [Phys. Rev. E {\bf 84}, 011125 
(2011)] leads to the
appearance of an unusual active asymmetric 
phase, in which the system sublattices 
are unequally populated. 
It differs from the  usual CP only by the fact that
 particles also interact 
with their next-nearest neighbor sites via a distinct strength 
creation rate and for the inclusion of an inhibition effect, proportional
to the local density.
Aimed at investigating the robustness of such asymmetric phase,
in this paper we  study the influence of
distinct interactions for two bidimensional CPs. 
In the first model, the interaction between first neighbors 
 requires a minimal neighborhood
of  adjacent particles for creating new offspring, whereas
 second neighbors interact as usual 
(e.g. at least one neighboring  particle is required).
The second model takes the opposite situation, in which the restrictive
dynamics is in the interaction between next-nearest neighbors sites.
Both models are investigated under mean field 
theory (MFT) and Monte Carlo simulations. 
In similarity with results by Martins {\it et. al.},
the inclusion of distinct sublattice interactions maintains
the occurrence of an asymmetric active phase and  reentrant 
transition lines. In contrast,
remarkable differences are presented, such as
 discontinuous  phase transitions (even between 
the active phases), the appearance 
of tricritical points and the stabilization of active phases under
larger values of control parameters. Finally, we have shown
that the critical behaviors are not altered due to the change
of interactions, in which the absorbing transitions belong
to the directed percolation (DP) universality class, whereas
second-order active phase transitions  belong to the Ising universality
class.

\end{abstract}

\maketitle

\section{Introduction}
Nonequilibrium phase transitions into absorbing states
have attracted considerable interest not only  for
the description of several
problems such as  wetting phenomena, spreading of diseases, 
chemical reactions \cite{c1,c2} but also for 
the search of experimental verifications \cite{c3}. In the most common
cases, phase transitions are second-order and belong to the directed
universality (DP) class \cite{c1}. However, the inclusion of distinct
dynamics (such as diffusion, disorder, laws of conservation, noise and 
others) not only may drastically change the phase transition and critical
behavior \cite{c2,henkel}, 
but also may exhibit new features such as Griffiths phases 
\cite{vojta}, formation of stable patterns \cite{vilar}, 
phase coexistence \cite{carlos} and others \cite{dic89}.
Recently,  Martins {\it et al.} \cite{dickman} have introduced a
two-dimensional  contact process (CP) \cite{cp} with  
sublattice symmetry breaking, in which
the dynamics is ruled by the competition between particle
creation at nearest and next-nearest neighbor occupied sites and
the annihilation also depends on the local particle density. 
Particles interact with their first- and second-neighbors
by means of a similar interaction rule, but the strengths of creation rates 
are different.
In addition to the usual absorbing and active (symmetric) 
phases, mean field theory (MFT) and Monte Carlo (MC)
analysis predict   
the appearance of an unusual active asymmetric phase,
in which in contrast to the symmetric phase 
the distinct sublattices are unequally populated. A phase transition, between
the symmetric and asymmetric phases is characterized by a spontaneous symmetry 
breaking.
All   absorbing phase transitions
belonging to the directed percolation (DP) \cite{c2} class, whereas the 
 transitions between active phases belong to
the Ising universality class.
Inspired by recent studies \cite{carlos,mario}, in which the  particle
creation   requiring a minimal neighborhood of occupied sites 
(instead of one particle as in the original CP) leads to the appearance of
a discontinuous absorbing phase transition, here
we give a further step   in the work by Martins {\it et al.} 
by including  such class of restrictive dynamics in order to raise three 
remarkable questions: 
First, does the competition between distinct sublattice interactions
(instead of only distinct creation rates) 
change the topology of the phase diagram? Is the asymmetric phase  maintained
by  changing the interaction rules? Are the classifications
of phase transitions altered?
To answer them, we analyze two distinct
models taking into account a minimum neighborhood of adjacent particles. 
Models are analyzed via mean-field
approximation and numerical simulations. Results have shown
 the  asymmetric phase  ``survives''
by the change of interactions but  pronounced changes 
in the phase diagram are found, such as  
discontinuous absorbing transitions, discontinuous transitions with
spontaneous breaking symmetry (instead of continuous transitions, as 
typically observed), the appearance of  of tricritical points, critical
end point and the extension of phases under
larger values of control parameters. \\This paper is organized
as follows: In Sec. II we describe the studied models and we 
show results under mean field analysis. In Sec. III we show numerical results
and we compare with those obtained in Sec. II. Conclusions
are done in Sec. IV.

\section{Models}
 
Let us consider a system of interacting particles placed on a 
square lattice of linear
size $L$ in which each site is empty or occupied by a particle. 
Dynamics is described as follows: 
Particles in a given sublattice $i$ (A or B) 
are created in empty sites with first- and second-neighbor 
transition rates  $\lambda_1n_{1i}/q$ and  $\lambda_2n_{2i}/q$  
respectively, being $\lambda_1$ and $\lambda_2$
the strength of creation  parameters, $n_{1i}$ and
$n_{2i}$ denote the number of particles in 
the first and second-neighbors
of the site $i$, respectively and $q$ is the
coordination number (reading  $4$ for a square lattice).   
In the model 1, the interaction between  first neighbors
is taken into account only if $n_{1i} \ge 2$, in such a way that
no contribution for the particle creation due to nearest neighbor
occupied sites occurs if  $n_{1i} \le 1$. 
In contrast, the interaction between second neighbors takes into
account $n_{2i} \ge 1$ particles, as in the usual CP. 
The model 2  is the opposite case,  in which  the transition rate 
between first neighbors  requires $n_{1i} \ge 1$ adjacent particles,
but the interaction between second neighbors contributes 
only  if $n_{2i} \ge 2$.  In order
to favor   unequal sublattice populations, 
a term  increasing with the number of nearest 
neighbors particles, in the form $\mu\,n_{1i}^2$,
is included in the annihilation rate \cite{dickman}. 
If $\mu=0$, one recovers the usual case in which
a particle is spontaneously annihilated with rate $1$.  

To characterize the phase transitions, the 
 sublattice particle densities $\rho_i$ ($i=A$ and $B$) are 
important quantities to measure. In the absorbing state  both sublattices
are empty, implying that
$\rho_A=\rho_B=0$. On the other hand, in an active symmetric ($as$) phase 
$\rho_A=\rho_B \neq 0$, whereas in an active asymmetric ($aa$) phase  
$\rho_A \neq \rho_B$ and $\rho=\frac{1}{2}(\rho_A+\rho_B) \neq 0$. Hence, in 
contrast to the $as$ phase, in the $aa$
phase the sublattices are unequally populated and the phase transition
is not ruled by the global density $\rho$, but for
 the difference of sublattice densities given
by $\phi=\frac{1}{2}(\rho_A-\rho_B)$. Unlike the $as$ phase,
in which $\phi=0$, in the $aa$ phase it follows that $\phi \neq 0$. 

\subsection{Transition rates and mean-field analysis}
\begin{figure}[h]
\centering
\includegraphics[scale=0.5]{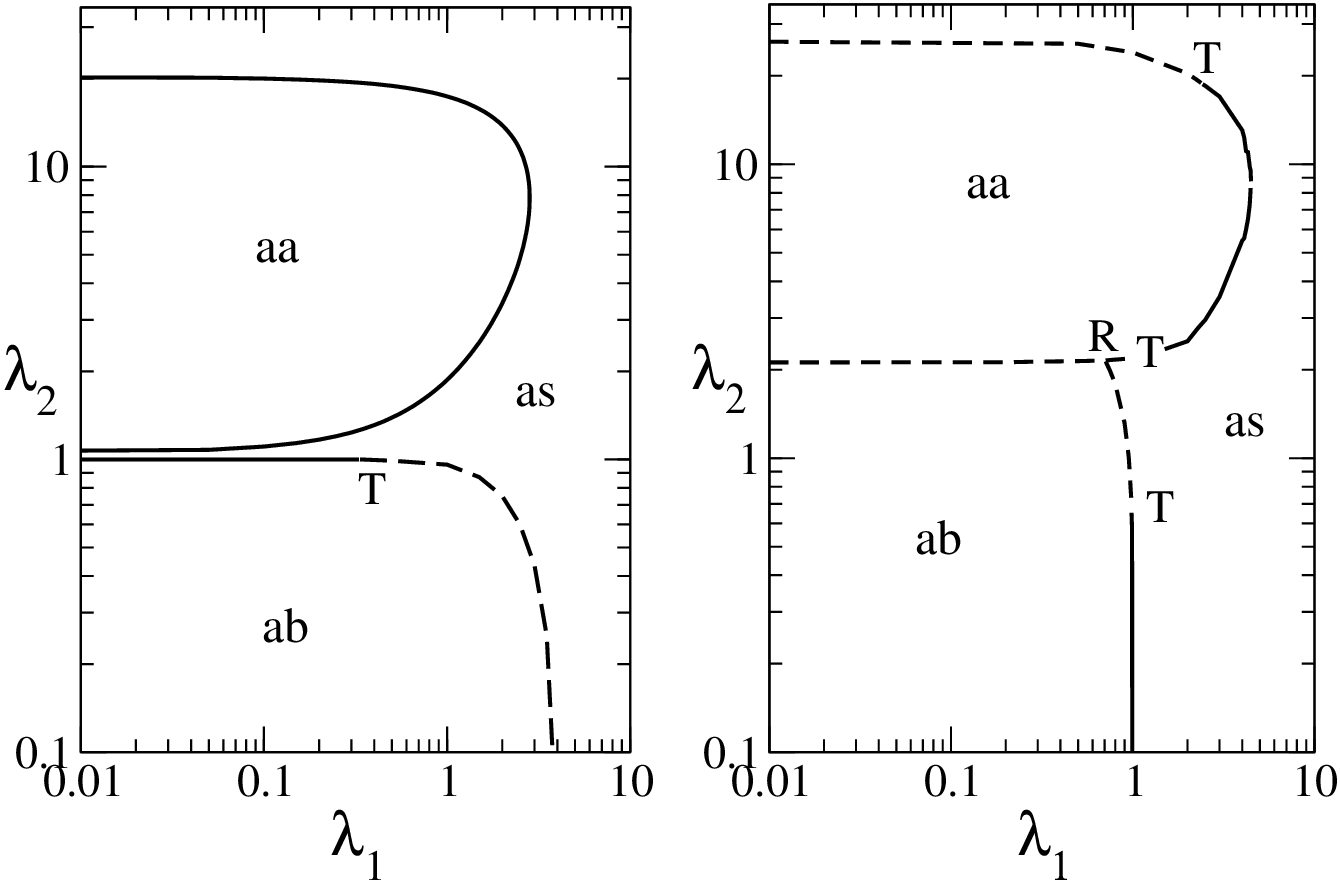}
\caption{Mean field phase diagram for 
the models 1 (left) and 2 (right) 
 for $\mu=1.0$. Absorbing, active-symmetric and 
active-asymmetric  phases well as  the triple and tricritical points 
are represented by the symbols $ab$, $as$, $aa$, $R$ and  $T$ respectively.}
\label{diagramtc}
\end{figure}

From the above model definitions, we can write down the time 
evolution of sublattice densities $\rho_A$ and $\rho_B$,  which
correspond to one site probabilities.
Let the symbols $\scalebox{0.75}{\fullcircle}$ and 
$\scalebox{1.02}{\fullsquare}$ 
to denote occupied sites belonging
to the sublattices $A$ and $B$, respectively. From the previous dynamic rules, 
it follows that  the time  evolution of $\rho_A$ and $\rho_B$ are given by
\[
\frac{d\rho_A}{dt}=\lambda_1[2P(\scalebox{1.02}{\fullsquare}\,
\scalebox{1.02}{\fullsquare}\,\scalebox{0.75}{\opencircle}
\scalebox{0.75}{\opensquare}\,\scalebox{0.75}{\opensquare})+
P(\scalebox{1.02}{\fullsquare}\,\scalebox{0.75}{\opensquare}\,
\scalebox{0.75}{\opencircle}\hspace{-0.07cm}\scalebox{1.02}{\fullsquare}
\,\scalebox{0.75}{\opensquare})+3P(\scalebox{1.06}{\fullsquare}\,
\scalebox{0.75}{\opensquare}\,\scalebox{0.75}{\opencircle}\hspace{-0.04cm}
\scalebox{1.02}{\fullsquare}\,\scalebox{1.02}{\fullsquare})+
P(\scalebox{1.02}{\fullsquare}\,\scalebox{1.02}{\fullsquare}\,
\scalebox{0.75}{\opencircle}\scalebox{1.02}{\fullsquare}\,\scalebox{1.02}
{\fullsquare})]+\]
\begin{equation}
+\lambda_2 P(\scalebox{0.75}{\opencircle}\scalebox{0.75}
\fullcircle)-[1+q^{2}\mu P(\hspace{-0.06cm}\scalebox{1.02}
{\fullsquare}\hspace{0.03cm})^2]
P(\hspace{-0.02cm}\scalebox{0.75}\fullcircle\hspace{-0.08cm})
\label{eq:ra1}
\end{equation}
\[
\frac{d\rho_B}{dt}=\lambda_1[2P(\scalebox{0.75}{\fullcircle}
\scalebox{0.75}{\fullcircle}\scalebox{0.75}{\opensquare}\,\scalebox{0.75}
{\opencircle}\scalebox{0.75}{\opencircle})+P(\scalebox{0.75}{\fullcircle}
\scalebox{0.75}{\opencircle}\scalebox{0.75}{\opensquare}\,
\scalebox{0.75}{\fullcircle}\scalebox{0.75}{\opencircle})
+3P(\scalebox{0.75}{\fullcircle}\scalebox{0.75}{\opencircle}
\scalebox{0.75}{\opensquare}\,\scalebox{0.75}{\fullcircle}\scalebox{0.75}
{\fullcircle})+P(\scalebox{0.75}{\fullcircle}\scalebox{0.75}{\fullcircle}
\scalebox{0.75}{\opensquare}\,\scalebox{0.75}{\fullcircle}\scalebox{0.75}
{\fullcircle})]+
\]
\begin{equation}
+\lambda_2 P(\scalebox{0.75}{\opensquare}\scalebox{1.02}
{\fullsquare})-[1+q^{2} \mu P(\hspace{-0.02cm}\scalebox{0.75}{\fullcircle}
\hspace{-0.08cm})^2]P(\hspace{-0.06cm}\scalebox{1.02}{\fullsquare}\hspace{0.03cm}).
\label{eq:rb1}
\end{equation}
for the model 1 and
\[
\frac{d\rho_A}{dt}=\lambda_1P(\scalebox{0.75}{\opencircle}\hspace{-0.05cm}
\scalebox{1.02}{\fullsquare}\hspace{0.06cm})+\lambda_2[2P(\scalebox{0.75}{\fullcircle}
\scalebox{0.75}{\fullcircle}\scalebox{0.75}{\opencircle}\scalebox{0.75}
{\opencircle}\scalebox{0.75}{\opencircle})+P(\scalebox{0.75}{\fullcircle}
\scalebox{0.75}{\opencircle}\scalebox{0.75}{\opencircle}\scalebox{0.75}
{\fullcircle}\scalebox{0.75}{\opencircle})+3P(\scalebox{0.75}{\fullcircle}
\scalebox{0.75}{\opencircle}\scalebox{0.75}{\opencircle}\scalebox{0.75}
{\fullcircle}\scalebox{0.75}{\fullcircle})+P(\scalebox{0.75}{\fullcircle}
\scalebox{0.75}{\fullcircle}\scalebox{0.75}{\opencircle}\scalebox{0.75}
{\fullcircle}\scalebox{0.75}{\fullcircle})]+
\]
\begin{equation}
-(1+q^{2}\mu\,P(\hspace{-0.06cm}
\scalebox{1.02}{\fullsquare}\hspace{0.03cm})^2)P(\hspace{-0.02cm}
\scalebox{0.75}{\fullcircle}\hspace{-0.08cm})
\label{eq:rav1}
\end{equation}
\[
\frac{d\rho_B}{dt}=\lambda_1P(\scalebox{0.75}{\opensquare}\,\scalebox{0.75}
{\fullcircle}\hspace{-0.05cm})+\lambda_2[2P(\scalebox{1.02}{\fullsquare}\,\scalebox{1.02}
{\fullsquare}\hspace{0.11cm}\scalebox{0.75}{\opensquare}\,\scalebox{0.75}{\opensquare}\,
\scalebox{0.75}{\opensquare})+P(\scalebox{1.02}{\fullsquare}\hspace{0.1cm}\scalebox{0.75}
{\opensquare}\hspace{0.1cm}\scalebox{0.75}{\opensquare}\hspace{0.05cm}\scalebox{1.02}
{\fullsquare}\hspace{0.1cm}
\scalebox{0.75}{\opensquare})+3P(\scalebox{1.02}{\fullsquare}\hspace{0.1cm}
\scalebox{0.75}{\opensquare}\,\scalebox{0.75}{\opensquare}\,
\scalebox{1.02}{\fullsquare}\,\scalebox{1.02}{\fullsquare})+
P(\scalebox{1.02}{\fullsquare}\,\scalebox{1.02}{\fullsquare}\hspace{0.14cm}
\scalebox{0.75}{\opensquare}\,\scalebox{1.02}{\fullsquare}\,
\scalebox{1.02}{\fullsquare})]+
\]
\begin{equation}
-(1+q^{2}\mu\,P(\hspace{-0.02cm}
\scalebox{0.75}{\fullcircle}\hspace{-0.08cm})^2)P(\hspace{-0.06cm}
\scalebox{1.02}{\fullsquare}\hspace{0.03cm})\,.
\label{eq:rav2}
\end{equation}
for the  model 2, where we are using the shorthand notations
 $\rho_A=P(\hspace{-0.01cm}\scalebox{0.75}\fullcircle\hspace{-0.08cm})$ and
 $\rho_B=P(\hspace{-0.04cm}\scalebox{1.02}{\fullsquare}\hspace{0.03cm})$.
Note that from
the above model definitions  it follows that 
 Eqs. (\ref{eq:ra1}) and (\ref{eq:rb1}) (model 1) and Eqs. 
(\ref{eq:rav1}) and (\ref{eq:rav2}) (model 2) are symmetric under 
$A \rightleftharpoons B$. In terms of the order parameter
$\phi$, the sublattice exchange implies that 
$\phi \rightleftharpoons  -\phi$.
Although the symmetric phase
remains unchanged under  $\rho_A \rightleftharpoons \rho_B$ 
(since $\phi=0$),  above symmetry is broken 
in the asymmetric phase (corresponding to 
$\phi^{*} \rightleftharpoons  -\phi^{*}$,
where $\phi^{*}$ is the steady value). Thus a spontaneous
symmetry breaking is expect to occur in the emergence of the
$aa$ phase.
The first inspection of the  phase diagrams can be achieved
by performing one-site mean field analysis. It consists of 
replacing a given $n-$site probability by a product of $n-$site
probabilities, 
in such a way that  Eqs. (\ref{eq:ra1}) and (\ref{eq:rb1}) become
\begin{equation}
\frac{d\rho_A}{dt}=\lambda_1\rho_B^2(1-\rho_A)[3-3\rho_B+\rho_B^2]
+\lambda_2(1-\rho_A)\rho_A-(1+q^{2}\mu\rho_B^2)\rho_A
\label{eq:ra2}
\end{equation}

\begin{equation}
\frac{d\rho_B}{dt}=\lambda_1\rho_A^2(1-\rho_B)[3-3\rho_A+\rho_A^2]
+\lambda_2(1-\rho_B)\rho_B-(1+q^{2}\mu\rho_A^2)\rho_B,
\label{eq:rb2}
\end{equation}
for the model 1 and Eqs. (\ref{eq:rav1}) and (\ref{eq:rav2}) become
\begin{equation}
\frac{d\rho_A}{dt}=\lambda_1(1-\rho_A)\rho_B+
\lambda_2\rho_A^2(1-\rho_A)[3-3\rho_A+\rho_A^2]-
(1+q^{2}\mu\,\rho_B^2)\rho_A
\label{eqE}
\end{equation}

\begin{equation}
\frac{d\rho_B}{dt}=\lambda_1(1-\rho_B)\rho_A
+\lambda_2\rho_B^2(1-\rho_B)[3-3\rho_B+\rho_B^2]-
(1+q^{2}\mu\,\rho_A^2)\rho_B\,.
\label{eqF}
\end{equation}
for the model 2. The steady solutions are obtained by taking 
 $\frac{d\rho_A}{dt}=\frac{d\rho_B}{dt}=0$ in both cases and
for a given set of parameters $\lambda_1, \lambda_2$ and $\mu$ 
we can obtain $\rho_A$ and $\rho_B$ by solving the system of two  coupled
equations, from which we have built the phase diagrams, as shown
in Fig. \ref{diagramtc}. From now on, we are going to refer to
$\phi$ only in terms of
its absolute value, calculated by  $\phi=\frac{1}{2}|\rho_A-\rho_B|$.

In conformity with results by Martins {\it et al.} \cite{dickman}, 
in which the $as$ phase
is not stable for $\mu=0$, we  have considered $\mu=1$ in all cases.
Such value is lower than considered in Ref. \cite{dickman}, in order
to exploit the role of distinct sublattice interactions in the $aa$ phase. 
In particular, for both models the system is constrained 
in the  $ab$  phase
for low $\lambda_1$ and $\lambda_2$, whereas
for  sufficient large $\lambda_2$ 
and $\lambda_1$ both 
$\rho_A$ and $\rho_B$ are close to $1$ and the system is in the $as$ phase.
The $aa$ phase is located for intermediate values of control
parameters and hence the phase diagrams
are reentrant.  For the model 1  the transition line, between
the $ab$ and $as$ phases  starts at 
$(\lambda_{1},\lambda_{2})=(0,1)$ and ends at $(4.1,0)$. It 
is second-order for low $\lambda_1$ but becomes 
discontinuous by increasing $\lambda_1$ with 
a tricritical point located at $(0.33,1)$.
In addition, the phase transitions between
active phases  (the $as-aa$ and $aa-as$ transition lines) are second-order,
starting at $(\lambda_1,\lambda_2)$=$(0,1)$ and $(0,20)$,
respectively and end at $(2.85,7.70)$. In summary, MFT 
 shows that the $aa$ phase is similar to that studied
in Ref. \cite{dickman} for the model 1 
and the inclusion of a distinct nearest-neighbor interaction
provoke qualitative changes  only  in the $ab-as$ transition line.

In contrast,  MFT  predicts
more substantial differences  for the model 2 than than
above mentioned results,  
as result of a distinct interaction between next-nearest neighbors.
No symmetric phase is presented for low $\lambda_1$, in such a way
that the  $ab-as$ and $as-aa$ transition lines (presented in
the model 1)  give rise to the $ab-aa$ coexistence line and
 meets the $ab-as$ and $as-aa$ lines in a triple point $R$ 
located at $(\lambda_1,\lambda_2)$=$(0.71,2.12)$. 
Also in contrast with above mentioned results, 
the $as-aa$ and $aa-as$ transition lines are first-order for low $\lambda_1$
and become continuous in tricritical points located at $(1.50,2.35)$ 
and $(2.43,18.7)$, respectively, giving
rise to correspondent critical lines.  Both 
critical lines meet at $(4.47,8.76)$. Finally,  
  the $aa$ phase extends  
for larger values of $\lambda_1$ and 
$\lambda_2$, but the $ab$ appears for
lower values of $\lambda_1$, as result of non restrictive
dynamics in the nearest-neighbor sites.    A tricritical
point located   at $(0.983,0.693)$ separates the coexistence
from the critical $ab-as$ transition lines.       
\section{Numerical results}
Numerical simulations have been performed for square lattices of
linear sizes
$L$ (ranging from $L=20$ to $80$) and periodic boundary conditions. For 
the  model 1, the actual MC dynamics is described as follows:
\begin{enumerate}
\item A particle $i$ is randomly selected from a list of currently
$N$ occupied sites.
\item The particle $i$ (for instance belonging
to the sublattice $A$) is annihilated with probability
$p_a=\frac{1+\mu n_B^2}{(1+\mu n_B^2+\lambda_1+\lambda_2)}$,
(being $n_B$ its number of nearest neighbor particles) 
and with complementary probability
$p_c=1-p_a$ the creation process is selected.
\item 
If the particle creation is performed, with probabilities 
$\lambda_1/(\lambda_1+\lambda_2)$ 
and $\lambda_2/(\lambda_1+\lambda_2)$ the first- and second-neighbor
particle interactions  will be chosen, respectively.
\item If the  first (second) neighbor interaction is chosen, 
one of its first (second) neighbors $j$ is randomly selected
and a particle will be created, provided  $j$ is empty and if 
at least two (one) of its first (second) neighbors 
are occupied.
\end{enumerate}
For the model 2 the last rule  
is replaced  in such a way that  the particle
is created in the site $j$ provided it is empty and at least 
one (two) of its first (second) neighbors
are occupied.

Numerical simulations have been improved by employing the quasi-stationary
method \cite{qs}. Briefly the method consists
of  storing a list of $M$ active configurations (typically one
stores $M=2000$ configurations)
and whenever the system falls into the absorbing state a
configuration is  randomly extracted from the list. 
The ensemble of 
stored configurations is continuously updated, where in practice,  
for each MC step a configuration belonging to the list is replaced with 
probability ${\tilde p}$ (typically one takes ${\tilde p}=0.01$)
by the actual system configuration, provided it is not absorbing.

Numerical simulations exhibit distinct behaviors
in the case of continuous and discontinuous transitions and
hence  distinct analysis are analyzed 
for characterize them. In the former case,  relevant thermodynamic 
quantities present algebraic behavior close to the critical point.
In particular, the order parameter $\phi$ and its variance
$\chi=\langle \phi^{2}\rangle-\langle \phi \rangle^{2}$ 
behaves as $\phi \sim (\lambda-\lambda_c)^{\beta}$ 
and $\chi \sim (\lambda-\lambda_c)^{-\gamma}$, respectively 
where $\beta$ and
$\gamma$ are associated critical exponents.   
Besides, at the critical point $\lambda_c$, $\phi$ and $\chi$ 
also exhibit power-law behaviors when simulated for finite
system sizes. According to the finite size scaling theory \cite{c1},
they behave as 
 $\phi \sim L^{-\beta/\nu_{\perp}}$ 
and $\chi \sim L^{\gamma/\nu_{\perp}}$, respectively where $\nu_{\perp}$ is the
critical exponent associated with the spacial length correlation. 
For the DP universality class in two dimensions, 
$\beta$, $\nu_{\perp}$ and $\beta/\nu_{\perp}$ read 
$0.5834(30)$, $0.7333(75)$ and $0.796(9)$, respectively, whereas for the Ising
universality class $\beta$, $\gamma$  and $\nu_{\perp}$  read
$1/8$, $7/4$ and $1$, respectively.

For locating the critical points, we study the  crossing among 
``cumulants'' curves. In particular,  a cumulant 
appropriate for absorbing transitions (being $\rho$ 
the order parameter) is the  moment ratio given by 
$U_2=\langle\rho^2\rangle/\langle\rho\rangle^2$. For DP transitions
in two dimensions,
it  assumes the universal value $U_{2c}=1.3257(5)$ at the critical point. 
In contrast, for the transition between active phases, a proper 
quantity to be studied is
fourth-order Binder cumulant \cite{binder}
\begin{equation}
U_4=1-\frac{\langle \phi^4 \rangle}{3 \langle \phi^2 \rangle}\,.
\end{equation} 
where in the present case $\phi$ is the difference between
sublattice densities, defined previously. The study
of above quantity is understood by recalling that the
$aa$ and $as$ phases are similar to the ferromagnetic and paramagnetic
ones found in the Ising model, respectively.
At the critical point, for systems
belonging to the Ising universality class, $U_4$ 
assumes the universal value $U_{4c}=0.61069...$ and hence
the crossing among distinct $L$'s will provide us an estimation
of the critical point.

In the case of first-order transitions, 
the probability order-parameter 
distribution  is an important quantity to characterize them, since in contrast
to second-order transitions,
it presents a bimodal shape at the phase coexistence.
Hence, a two peak probability distribution for larger system
sizes will be used as the indicator of a phase coexistence.

\begin{figure}[h]
\centering
\includegraphics[trim=0.001cm 0.001cm 0.001cm 0.1cm, clip=true, scale=0.4]{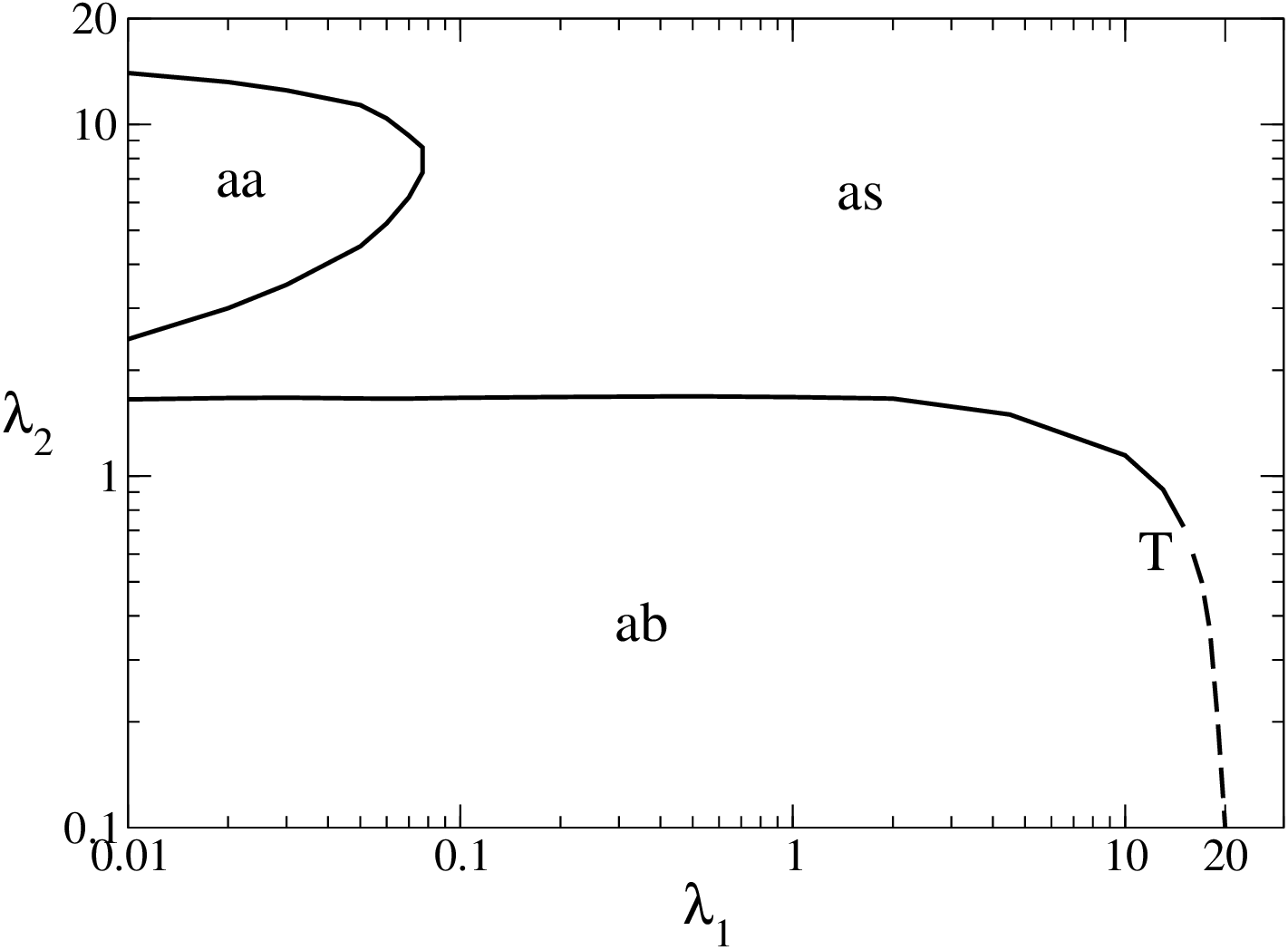}
\caption{For the model 1, phase diagram in the
 $\lambda_1-\lambda_2$ space obtained from MC  simulations.
Dashed and continuous line
denote discontinuous and continuous phase transitions,
respectively. The symbols $ab$, $as$, $aa$  and $T$ denote
the absorbing, active symmetric and active asymmetric
phases and a tricritical point, respectively.}
\label{diagramtce}
\end{figure}

After presenting the methodology, let us show numerical results and 
comparing with the MFT ones.
In Fig. \ref{diagramtce} we show the phase diagram obtained
from numerical simulations for model 1. The topology of the 
phase diagram is similar
to that obtained from MFT, including the existence
of absorbing, symmetric and asymmetric phases and the following
 $ab-as$, $as-aa$ and $aa-as$  transition lines. 
Also in similarity with MFT, the $ab-ss$ line
is continuous for low  $\lambda_1$ and becomes
discontinuous by increasing such nearest-neighbor creation parameter,
whereas  the phase transition between active phases are second-order. However, 
differences with the MFT are observed. In particular, 
in similarity with the results  by Martins {\it et al.} 
in Ref. \cite{dickman},  the $aa$ phase is placed for lower 
values of control parameters than those obtained from the MFT. In contrast, 
the $ab-as$ transition line extends for relatively
larger $\lambda_1$'s.

\begin{figure}[h]
\centering
\includegraphics[scale=0.34]{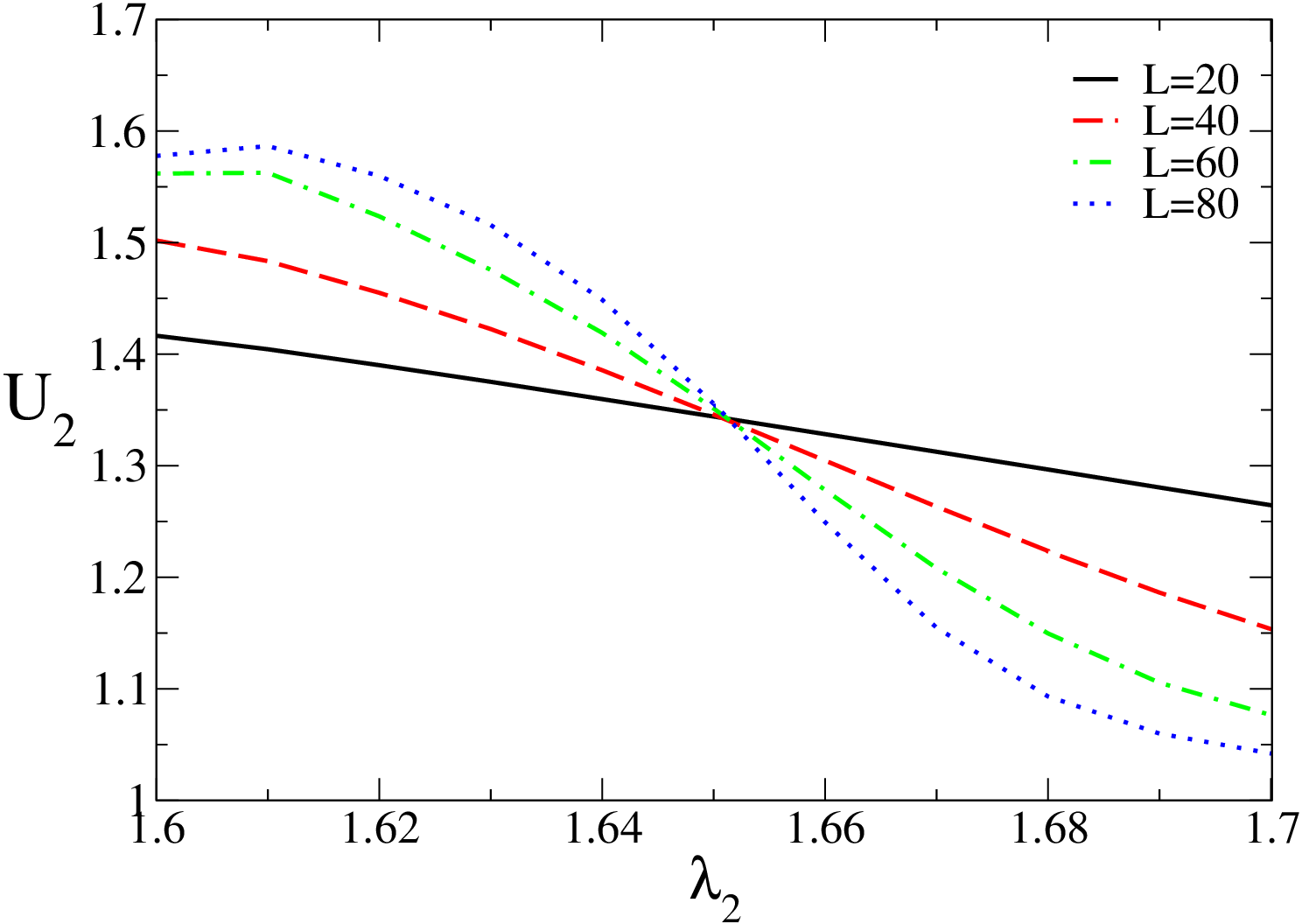}
\caption{Moment ratio $U_2$ versus $\lambda_2$ for distinct L`s and $\lambda_1=0.01$.}
\label{fig:u2}
\end{figure}

\begin{figure}[h]
\centering
\includegraphics[scale=0.45]{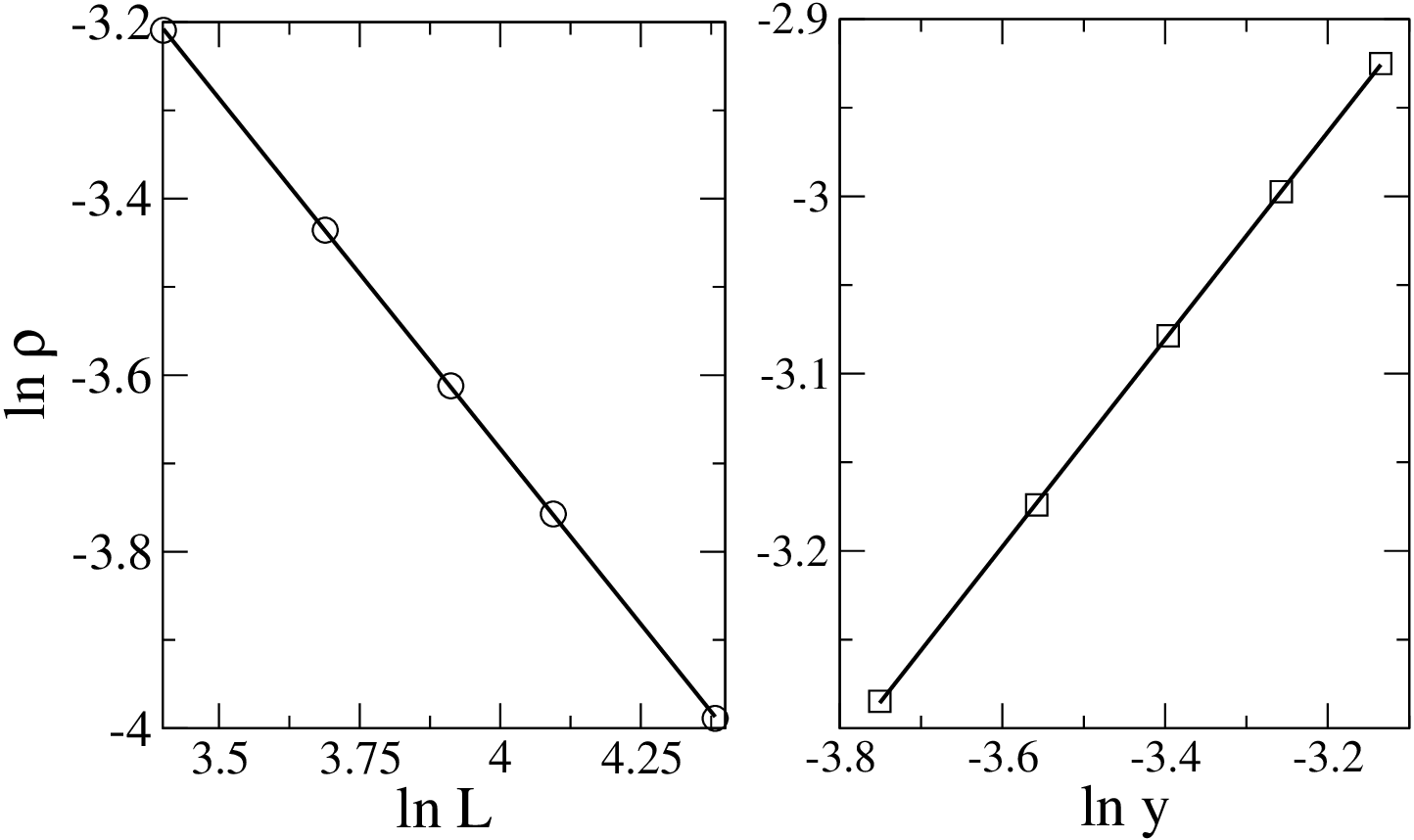}
\caption{In the left, log-log plot of 
$\rho$ vs $L$  for $\lambda_1=0.01$
and  $\lambda_{2c}=1.6515$. In the right, 
log-log plot of $\rho$ vs
$y=\lambda_2-\lambda_{2c}$ for  $\lambda_1=0.01$ and $L=80$.
The left and right curves have slopes $\beta/\nu_{\perp}=0.796(9)$
and $\beta=0.5834(30)$, respectively.}
\label{fig:beta_nu}
\end{figure}

After describing the main features of the phase diagram, let us 
show some explicit results for 
distinct points of the phase diagram. Starting
from the $ab-as$ transition line, in Fig. \ref{fig:u2} we plot
the moment ratio $U_2$ for distinct system sizes and $\lambda_1=0.01$. 
Note that all curves cross at $\lambda_{2c}=1.6515(5)$ with 
$U_2=1.34(2)$, which is close to the 
universal DP value $1.3257(5)$.
In fact,
as shown in Fig. \ref{fig:beta_nu}, for $\lambda_{2c}=1.6515$
we find the critical exponents $\beta/\nu_{\perp}=0.794(2)$
and $\beta=0.584(2)$, which are compatible with the DP values 
(solid lines).
Results for  other critical points (not shown) confirm
that the second-order transitions between the $ab$ and $as$ phases
belong to the DP universality class.

\begin{figure}[h]
\centering
\includegraphics[scale=0.34]{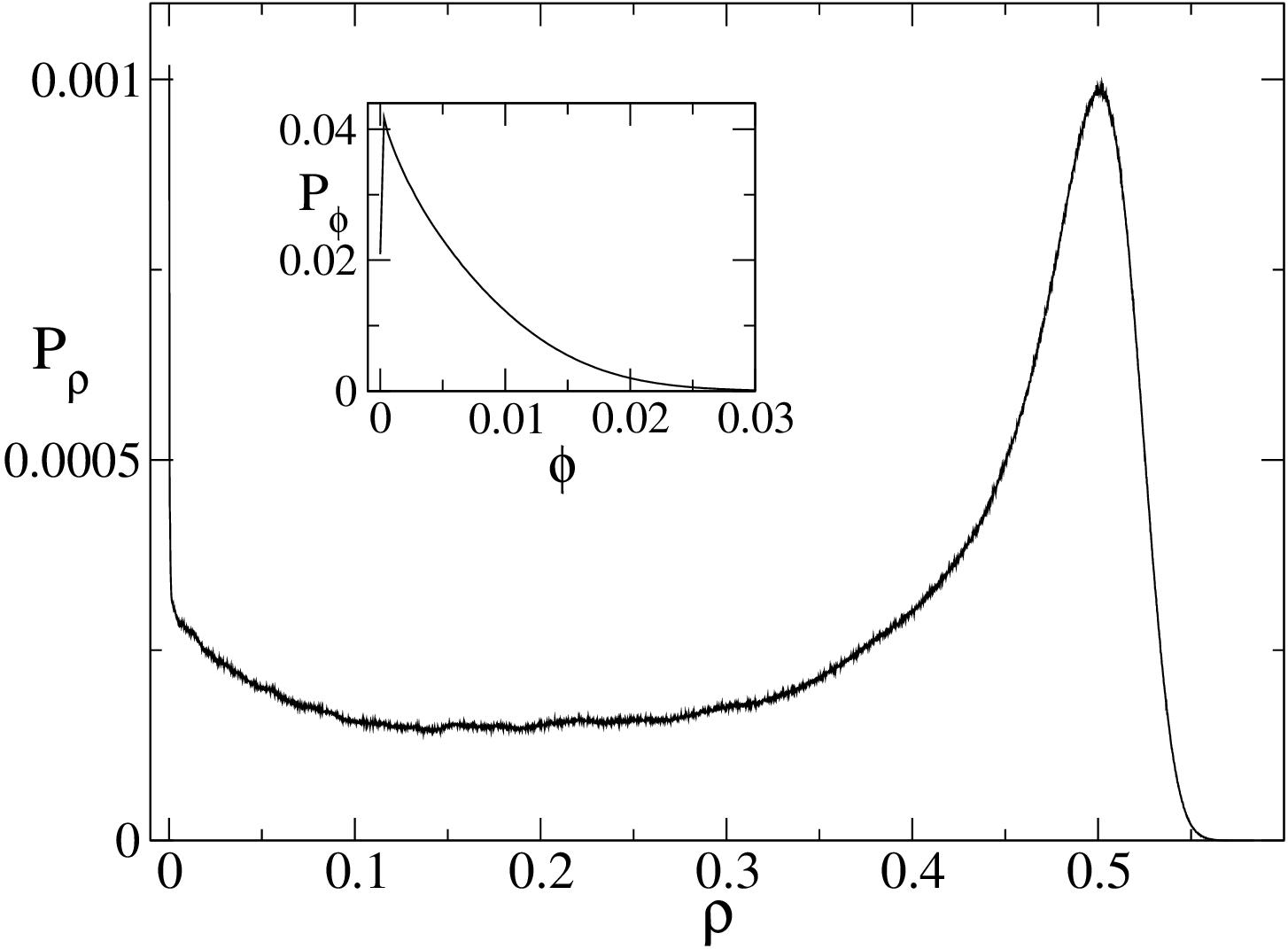}
\caption{For the model 1, the quasi-stationary probability distribution 
$P_{\rho}$   for $\lambda_1=18$ and $\lambda_2=0.3618$ and  $L=80$. 
In the inset, $P_{\phi}$ for the system density $\phi$.}
\label{fig:L2}
\end{figure}

In Fig. \ref{fig:L2} we show results for $\lambda_1=18$.
For $L=80$ and  $\lambda_2=0.3618$, the probability
distribution $P_{\rho}$ presents
two equal peaks at   distinct densities ($\rho=0.0002$ and $0.501$)
and together a single peak centered at $\phi \sim 0$ for $P_{\phi}$, 
the phase transition between the 
$ab$ and $as$ phases for $\lambda_1=18$ is first-order. 

\begin{figure}[h]
\centering
\includegraphics[scale=0.4]{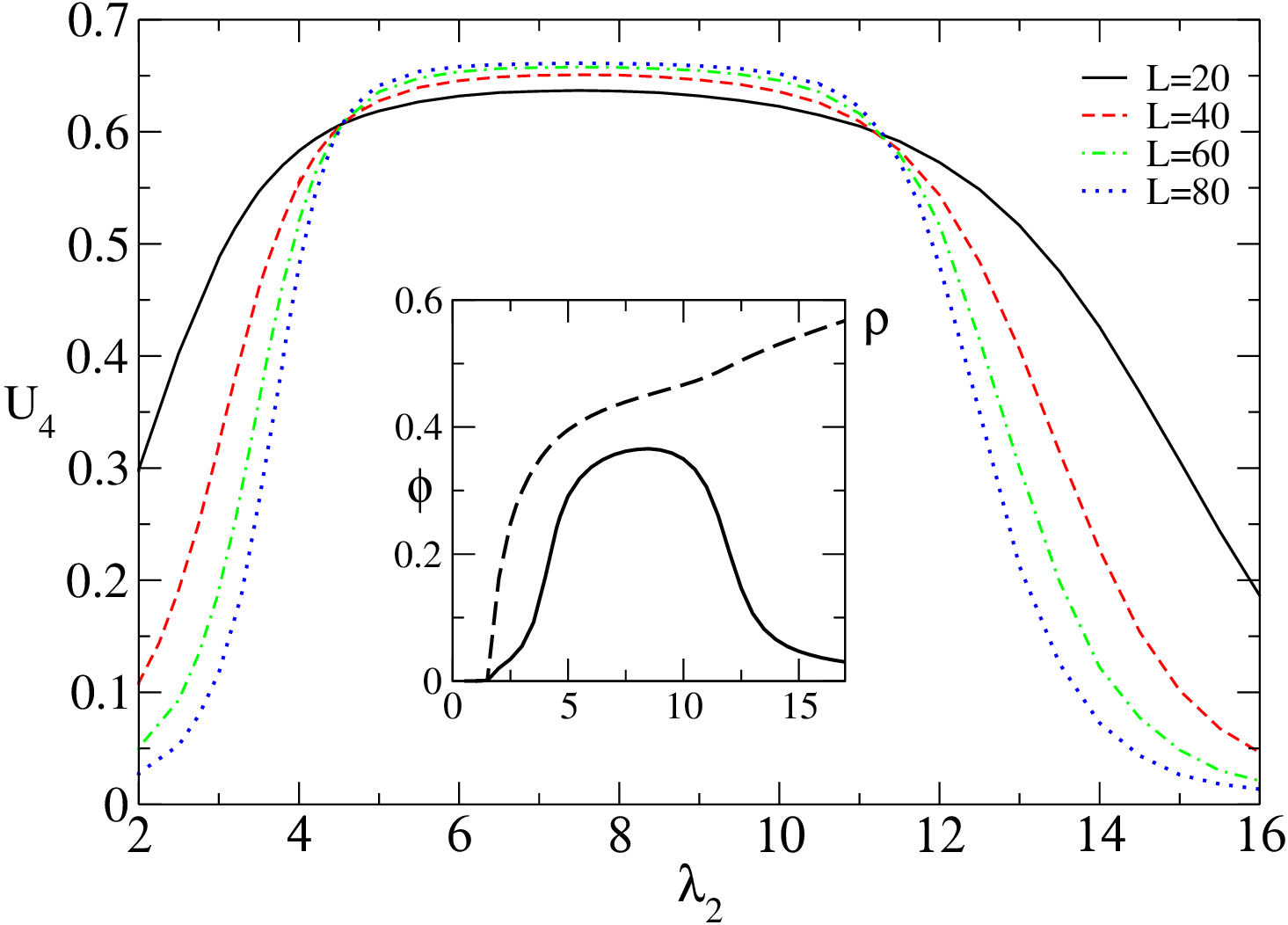}
\caption{Reduced fourth order cumulant $U_4$ versus $\lambda_2$ 
for distinct $L$'s and $\lambda_1=0.05$. In the inset, we plot
the order-parameter (continuous lines) and the system density
$\rho$ (dotted) vs $\lambda_2$ for $L=80$.}
\label{fig:binder}
\end{figure}

\begin{figure}[h]
\centering
\includegraphics[trim=0.001cm 0.001cm 0.001cm 0.05cm, clip=true,scale=0.4]{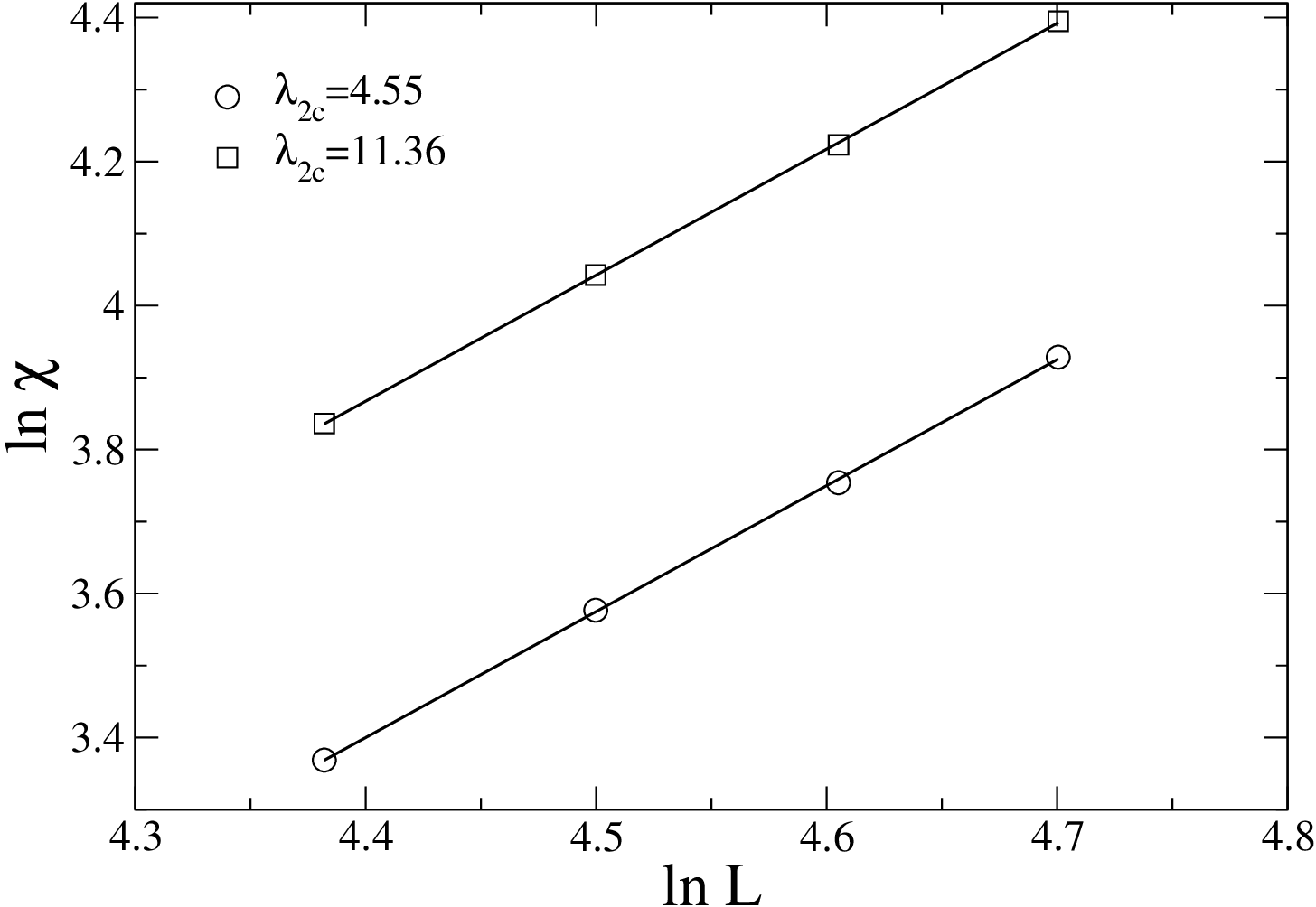}
\caption{Log-log plot of order-parameter variance $\chi$
 versus $L$ for $as-aa$ (circles) and
$aa-as$ (squares) for $\lambda_1=0.05$.
The straight lines have slopes $7/4$. }
\label{fig:ising}
\end{figure}

Next,  we study the phase transition between active phases,
whose  results are exemplified for $\lambda_1=0.05$ and 
shown in Fig. \ref{fig:binder}. Note that for 
$2.5<\lambda_2<5.0$, $\phi$ has a sharp increase followed
by a less pronounced 
 change of $\rho \neq 0$ (inset of Fig. \ref{fig:binder}),  signaling
the emergence of the  $as-aa$ phase transition. Results
for  $U_4$ show that all curves 
(for distinct system sizes) cross at $\lambda_{2c}=4.55(5)$ with  
$U_{4}=0.605(5)$, which is very close to the universal value $U_4=0.61069$ and
highlighting that such second-order 
phase transition belongs to the Ising universality class \cite{dickman}.
By increasing further $\lambda_2$, $\phi$ reaches a maximum and 
starting decreasing until vanishing. Such sharp behavior, accompanied
by a smooth variation of $\rho$, are consistent to the $aa-as$ phase transition.
We see that all cumulant curves cross  
 at $\lambda_{2c}=11.36(5)$ with $U_{4c}=0.60(1)$, 
which is also consistent  with the Ising value.  Note that
in the $aa$ phase $U_4 \rightarrow 2/3$ by increasing the system $L$,
signaling that the spontaneous symmetry breaking is similar
to that found in the Ising model.
To confirm above expectations, we analyze the order-parameter variance 
$\chi$ for finite system sizes, whose results 
are shown in  Fig. \ref{fig:ising}. At above critical points, 
we found the exponents
$\gamma/\nu=1.75(1)$ and $1.75(1)$, which are in good accordance with the value
$7/4$ and  hence confirming above expectations.

\begin{figure}[h]
\centering
\includegraphics[scale=0.4]{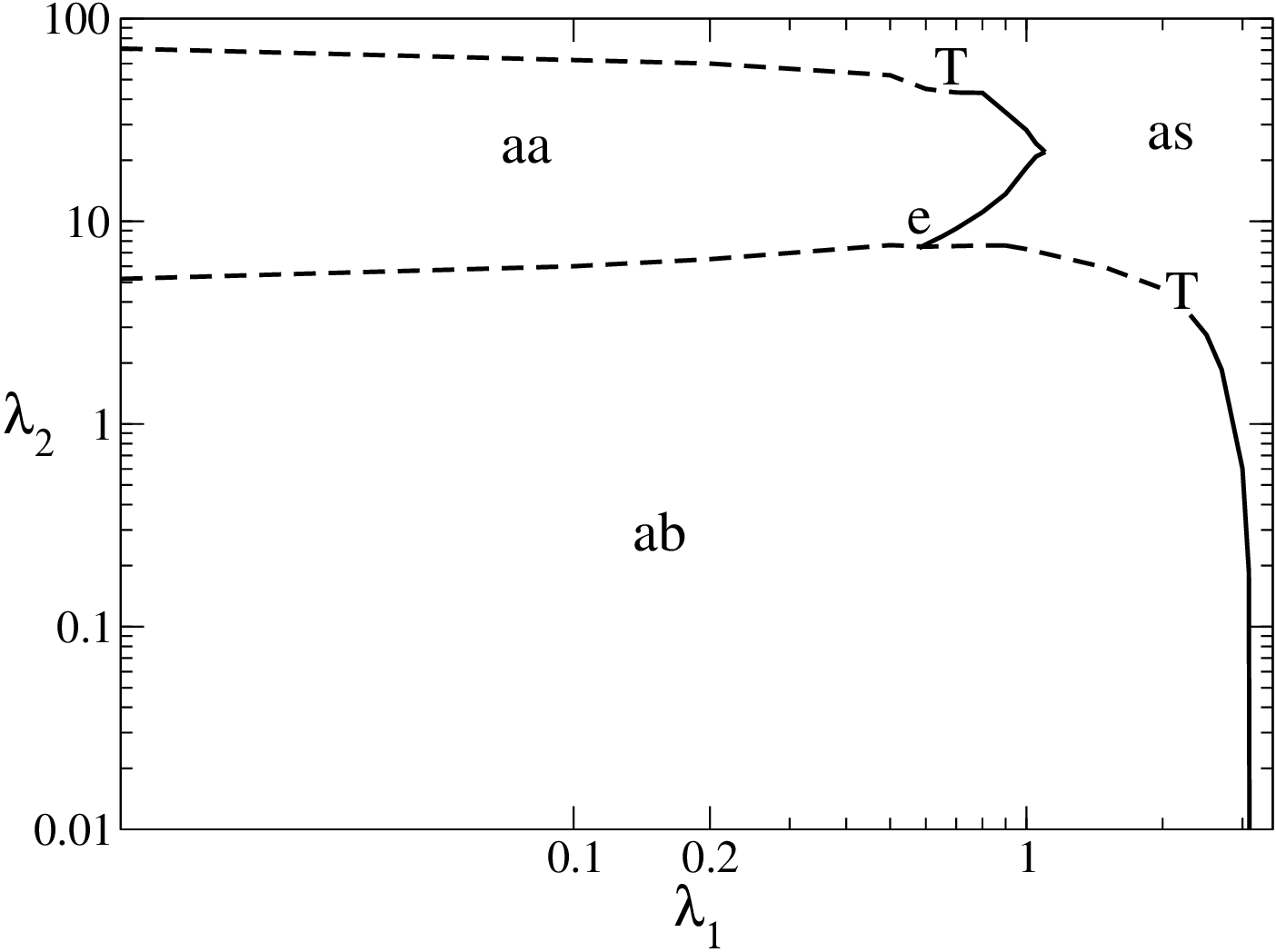}
\caption{For the model 2, the phase diagram 
in the plane  $\lambda_1-\lambda_2$, obtained
from MC simulations. Dashed and continuous line
denote discontinuous and continuous phase transitions,
respectively. The symbols $ab$, $as$, $aa$, $e$ $T$ denote
the absorbing, active-symmetric and active-asymmetric
phases, critical end point and  tricritical points, respectively.}
\label{diagramtc2}
\end{figure}

The phase diagram for the model 2 is shown in Fig. \ref{diagramtc2}.
In similarity with the model 1 and results by Martins
{\it et al.}, the
inclusion of  restrictive interaction 
between next nearest neighbor sites also maintains the $aa$
phase for intermediate values of $\lambda_2$. However,
confirming some MFT expectations,  there are more
pronounced differences with respect to above mentioned
results.
More specifically, the phase $as$  exists solely  to 
larger values of $\lambda_1$,  in such a way that no $ab-as$ 
transition line is presented for 
low $\lambda_1$. Besides,  the $aa$ phase is constrained
by transition lines that are first-order 
and become critical by increasing $\lambda_1$. Hence, in contrast
with above mentioned results, the symmetry breaking  occurs through
a discontinuous phase transition for low $\lambda_1$. Also unlike
previous cases,  tricritical
points separate the $as-aa$ and $aa-as$ coexistence lines from those
respective critical curves.
As a result of restrictive interaction between
next-nearest neighbor particles, the $aa$ phase extends
for very larger values of control parameters than model 
1 and those from Ref. \cite{dickman}. 
Also confirming the MFT expectations, 
the  phase transition between $ab$ and $as$ 
phases is critical and become discontinuous by lowering $\lambda_1$. 
Despite  above similarities, remarkable differences 
with MFT results are presented. There is no triple point 
in which $ab$, $as$ and $aa$ phases coexist. Instead, 
the critical $as-aa$ line meets the coexistence line $ab-aa$ 
in a critical end point $(e)$ (located at 
$[\lambda_1,\lambda_2]=[0.58(1),7.37(1)]$), giving
rise to the $ab-as$ phase coexistence. 
Besides, the $aa$ phase extends for much
 larger $\lambda_2$ and  lower $\lambda_1$ than
those obtained from MFT, but
the critical line $ab-as$ extends for larger values of $\lambda_1$
than  MFT predictions.

\begin{figure}[h]
\centering
\includegraphics[scale=0.5]{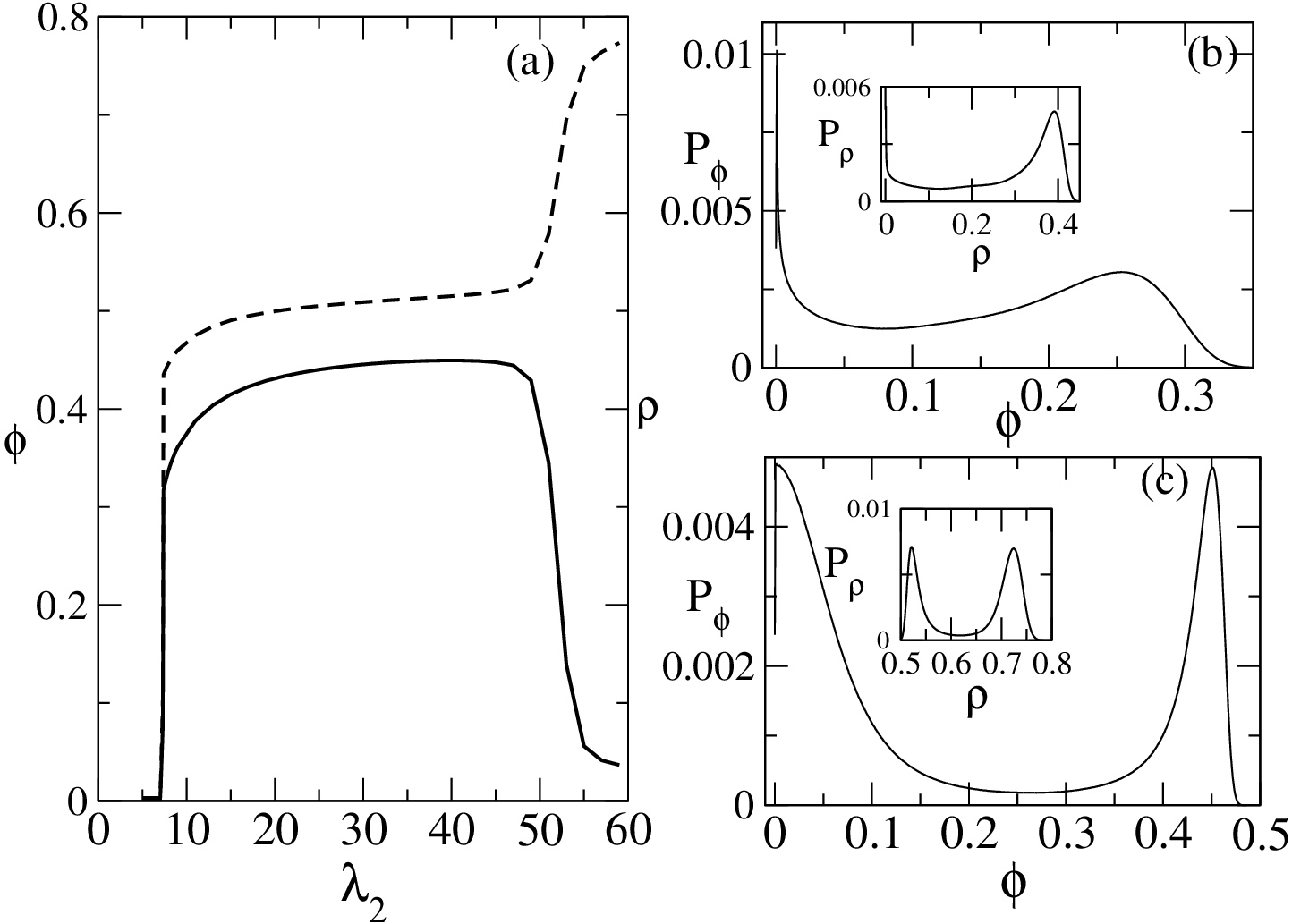}
\caption{For $L=80$ and $\lambda_1=0.5$, in $(a)$ the order parameter $\phi$ 
(continuous lines) and the system density $\rho$ (dashed lines) vs $\lambda_2$. 
In $(b)$ and $(c)$ probability distributions $P_{\phi}$ and $P_{\rho}$
(inset)   for  $\lambda_2=6.39$ and $\lambda_2=50.96$, respectively. }
\label{figx}
\end{figure}

In order to exemplify all above features
of the phase diagram, now we show  explicit results for 
distinct  points of the phase diagram. Starting from
 the $ab-aa$ and $aa-as$ coexisting phases, in Fig. \ref{figx}
we show explicit results for $\lambda_1=0.5$.
For low $\lambda_2$ the system is constrained in the $ab$ phase and
at a threshold value ($\lambda_2 \sim 6.39(1)$ for $\lambda_1=0.5$), both $\rho$
and $\phi$ changes abruptly, signaling the $ab-aa$ phase coexistence.
As for the model 1, in  the
$aa$ phase $\rho$ presents a smooth variation, implying that
the change of $\phi$ as $\lambda_2$ increases comes  
mainly from the spatial  redistribution of particles in sublattices. 
 In addition, the $aa$ phase extends for 
expressively larger values of $\lambda_2$.
Probability distributions in Fig. \ref{figx}$(b)$
reinforce the $ab-aa$ 
phase transition to be first-order, with two peaks (centered at 
$\phi \sim 0$ and $\phi \sim 0.25$  for $P_{\phi}$ 
and $\rho \sim 0$ and $\rho \sim 0.4$ for $P_{\rho}$),
 in consistency with  the observed jumps.
At a second threshold value ($\lambda_2=50.96(1)$ 
for $\lambda_1=0.5$) $\phi$ vanishes
abruptly (with $\rho$ presenting 
a certain increase), signaling the $aa-as$ phase transition.  Once again,
probability distributions in Fig. \ref{figx}$(c)$
confirm such transition to be first-order, with two peaks centered at 
$\phi \sim 0$ and $\phi \sim 0.45$ ($\rho \sim 0.52$ and $\rho \sim 0.73$)
for $P_{\phi}$ ($P_{\rho}$).

\begin{figure}[h]
\centering
\includegraphics[scale=0.5]{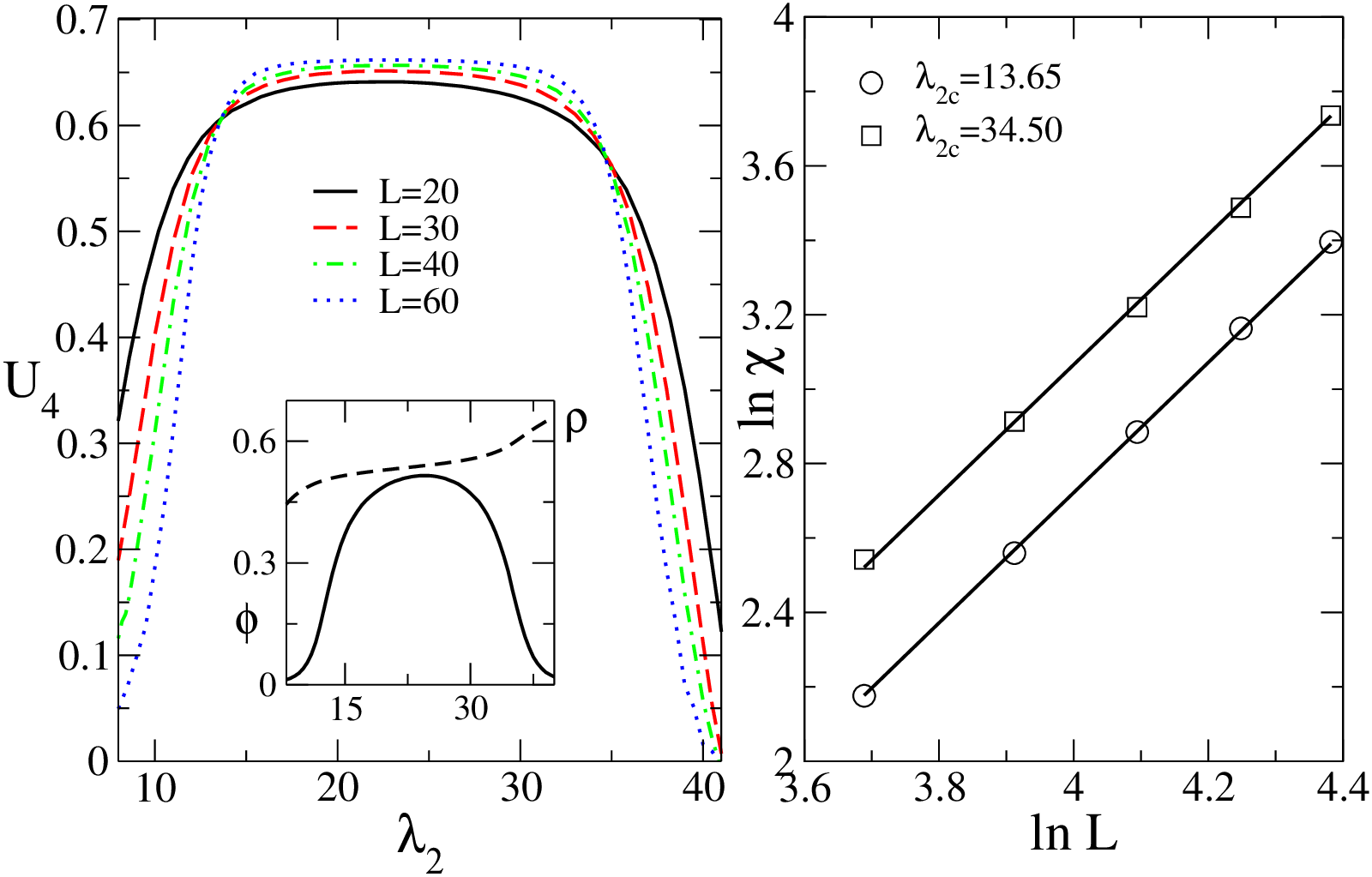}
\caption{In the left, $U_4$ vs $\lambda_2$ for distinct
system sizes for $\lambda_1=0.9$. In the inset, black
and dashed lines  show the order parameter
$\phi$ and system density $\rho$ vs $\lambda_2$ for $L=80$, respectively.
In the right, log-log plot of $\chi$ vs $L$ at the
$as-aa$ (circles) and $aa-as$ (squares) critical points. The straight
lines have slopes $7/4$.}
\label{cum2thr2v}
\end{figure}

Similar above behaviors are verified for other values of $\lambda_2$. 
Numerical results show that  $aa-as$ transition
lines become critical at $(\lambda_1,\lambda_2)$= $(0.65(3),44(1))$.  
In Fig. \ref{cum2thr2v} we show results for $\lambda_1=0.9$,
in order to exemplify the  second-order transition between active phases. 
In the interval
$10<\lambda_2<15$  $\phi$ increases 
substantially followed by small variation of $\rho \neq 0$. The first
crossing curves for $U_4$
occurs at $13.65(5)$ with  $U_4=0.61(1)$, which is consistent
with a second-order  Ising phase transition.
The maximum value of  $\phi$ (for $\lambda_1=0.9$) 
yields at $\lambda_2 \sim 25$, from which 
$\phi$ starts decreasing until vanishing and
no pronounced changes of $\rho$, signals
the  $aa-as$ phase transition.
For such transition, all reduced cumulant $U_4$  curves
cross in the interval  $34.5(2)$ with $U_4=0.59(2)$-also consistent
with previous phase transitions.  
By measuring the critical exponents, we obtain in both cases
values consistent with the value $7/4$, in similarity
with previous results. 

\begin{figure}[h]
\centering
\includegraphics[scale=0.35]{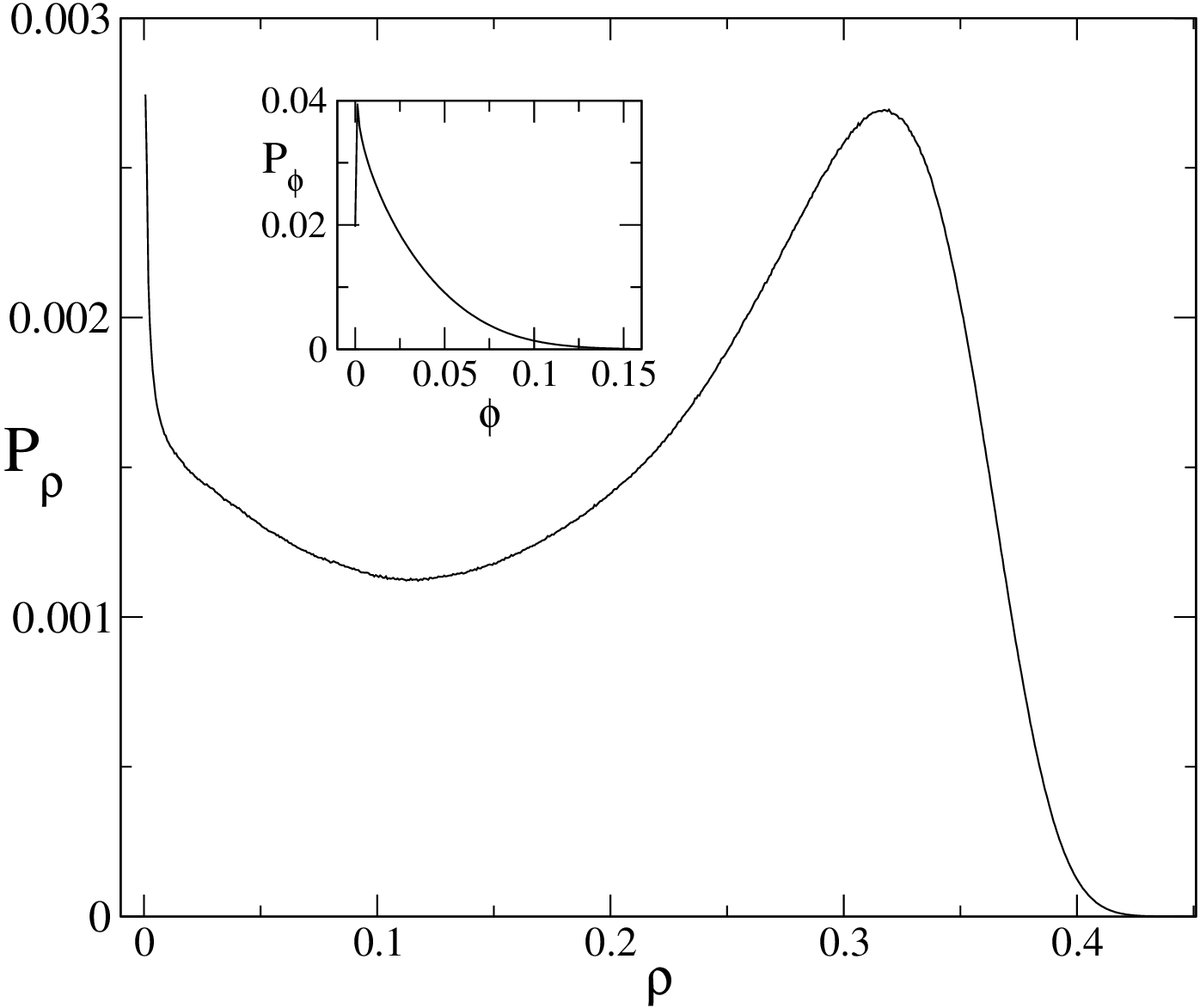}
\caption{Quasi-stationary order-parameter probability distribution $P_{\rho}$  
for $\lambda_1=1.5$ and $L=80$. In the inset,
the $P_{\phi}$ for the order-parameter $\phi$.}
\label{hist2}
\end{figure}

In the last analysis, we examine the transition between
the absorbing and active (symmetric) phases, whose
results are exemplified in Fig. \ref{hist2} 
for $\lambda_1=1.5$. The probability distribution
$P_{\rho}$   has equal height peaks 
centered at densities $\rho \sim 0$ and $\rho \sim 0.317$,
and together the single peak of $P_{\phi}$  centered at $\phi \sim 0$, such
result confirms the $ab-as$ phase coexistence for $\lambda_1=1.5$.

Finally, we plot results for $\lambda_1=3$, in order
to exemplify the critical $ab-as$ transition.  
As shown previously for the model 1, all $U_2$ curves cross at the 
point  $\lambda_{2c}=0.602(2)$ with $U_2=1.34(1)$, 
which is close to the DP value $1.3257(5)$. 
At the above crossing point,  $\rho$  behaves algebraically with an exponent
consistent with the DP value $\beta/\nu_{\perp}=0.796(9)$, illustrating
that the critical $ab-as$ line belongs to the DP universality class.
\begin{figure}[h]
\centering
\includegraphics[scale=0.45]{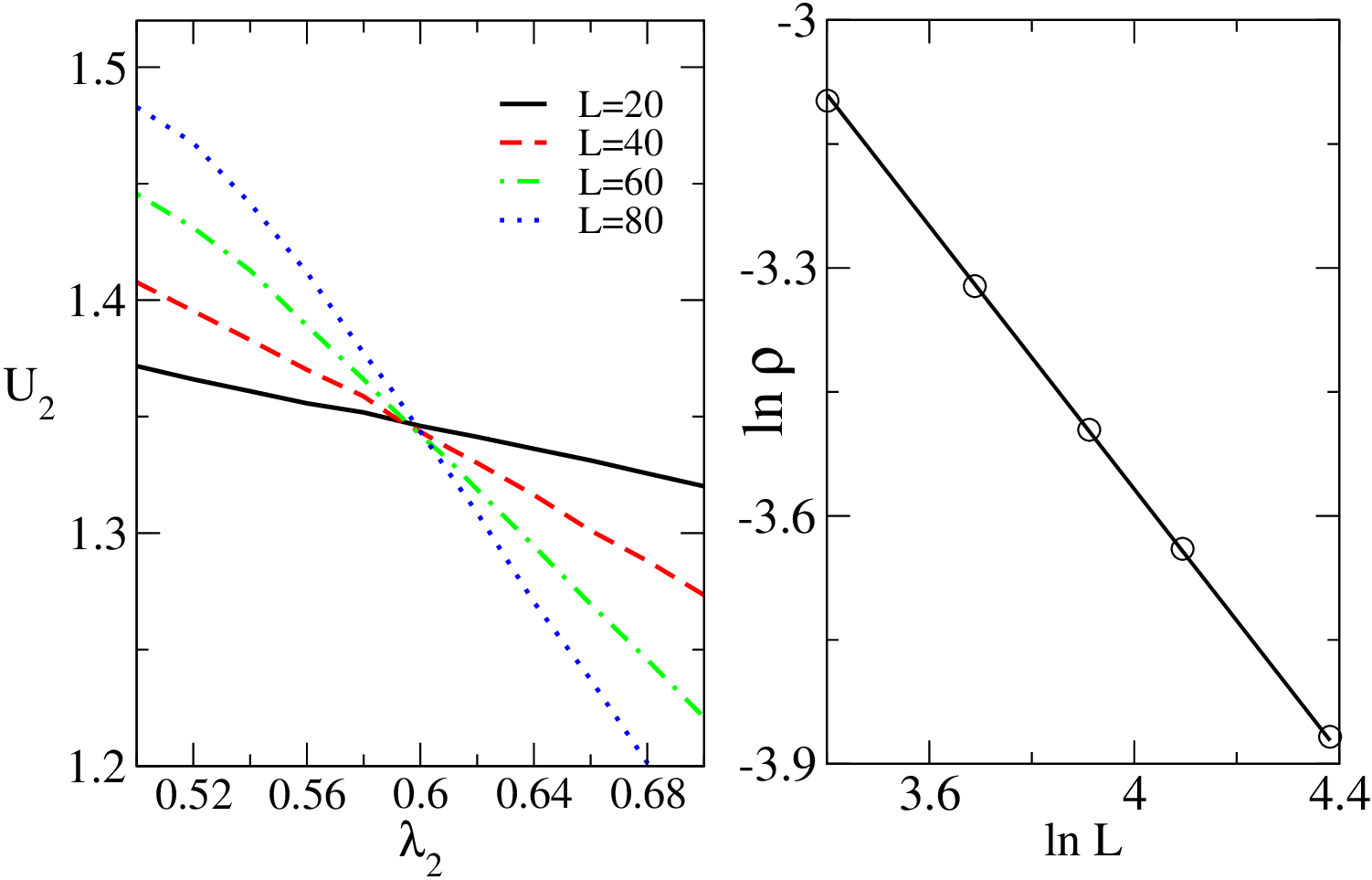}
\caption{In the left, moment ratio $U_2$ versus $\lambda_2$ for  
$\lambda_1=3$ and distinct system sizes. In the right,  
log-log plot of $\rho$ vs $L$ at the  
transition point $\lambda_{2c}=0.602$. The  straight line has 
slope $\beta/\nu_{\perp}=0.796(9)$.}
\label{figab2}
\end{figure}

\section{Conclusions}
The original two-dimensional contact process with creation at both
nearest and next-nearest neighbors and particle suppression  exhibit
a novel phase structure presenting
a continuous phase transition with 
spontaneous broken-symmetry phase and sublattice ordering \cite{dickman}.
Aimed at exploiting the robustness of such asymmetric phase
and the possibility of  distinct  phase transitions, in
this paper we studied the effect of distinct sublattice interactions 
(instead of only distinct creation rates as in the original model). 
Two distinct models were considered. In both cases,
results  confirm that the competition between
first and second-neighbor creation rates and particle suppression
are  fundamental requirements for the presence of an asymmetric active phase.
In addition, the inclusion of distinct competing interactions lead to novel
phase structures, summarized as follows: A restrictive interaction
between nearest neighbor sites (model 1) changes
the absorbing phase transition (in contrast with the original model), but
not the asymmetric phase. 
More pronounced changes are found by taking
the restrictive interaction between second-neighbor particles 
(model 2). It
not only prolongs greatly the asymmetric phase under larger values of
control parameters but also shift the phase transitions,
from continuous to  discontinuous, even between 
the active phases. This latter result is particularly interesting since
it reinforces the  role of restrictive interactions 
as a minimal mechanism 
for the appearance of first-order phase transitions \cite{carlos}. Initially
studied for  absorbing phase transitions, our results revealed that this
ingredient  is more general, changing the nature of  
distinct phases structures. Although
predicted by the mean field theory, it is worth mentioning that discontinuous
absorbing transitions under the studied restrictive interactions do not occur
in one-dimensional systems  \cite{hinri00}. 
The resemblance between $as-aa$ and ferromagnetic-paramagnetic
Ising model transition also provides a reasoning why 
such transitions can not occur in one-dimension.
In fact, 
results obtained by Martins {\it et al.} for the original version confirm this.
As a final remark, we note that possible
extensions of the present work includes exploiting the   
influence of distinct dynamics (such as diffusion,  annihilation
rules) in the asymmetric phase. This should be addressed
in a ongoing work.

\section{Acknowledgments}

The authors wish to thank Brazilian scientific agency CNPq, INCT-FCx 
for the financial support
and Universidade Federal do Paran\'a (UFPR) for providing basic
infrastructure to conduct the work. Salete Pianegonda also wishes to thank 
the Physics Department of the Federal Technological University, 
Paran\'a (DAFIS-CT-UTFPR)
for providing the access to its high-performance computing facility.

\end{document}